\documentstyle[12pt,aasms]{article}

\tighten

\newcommand{\beq}{\begin{equation}}
\newcommand{\eeq}{\end{equation}}
\newcommand{\etal}{{et al.\ }}

\newcommand{\sfrac}[2]{{\textstyle\frac{#1}{#2}\,}}

\lefthead{KARAS}
\righthead{LIGHT-CURVE FITTING-FORMULA}
\slugcomment{Scheduled for The Astrophysical Journal, Vol. 470 (October
20, 1996)}

\begin{document}
\title{LIGHT CURVE OF A SOURCE ORBITING AROUND A~BLACK HOLE:
A~FITTING-FORMULA}

\author{V. Karas\altaffilmark{1}}
\affil{Astronomical Institute, Charles University, \v{S}v\'edsk\'a 8,
CZ-150\,00 Prague, Czech Republic; karas@mbox.cesnet.cz}

\altaffiltext{1}{Also at Scuola Internazionale Superiore di Studi
Avanzati, Trieste; Department of Astronomy and Astrophysics, G\"oteborg
University and Chalmers University of Technology, G\"oteborg}

\begin{abstract}
A simple, analytical fitting-formula for a photometric
light curve of a source of light orbiting around a black hole is
presented. The formula is applicable for sources on a circular orbit
with radius smaller than 45 gravitational radii from the black hole.
This range of radii requires gravitational focusation of light rays and
the Doppler effect to be taken into account with care. The
fitting-formula is therefore useful for modelling the X-ray variability
of inner regions in active galactic nuclei.
\end{abstract}

\keywords{galaxies: active --- accretion, accretion disks --- X-rays:
 galaxies --- black hole physics}

\section{INTRODUCTION}
In their seminal paper, Cunningham \& Bardeen
(1973) studied photometric light curves of a point-like source of light
orbiting in the equatorial plane of a rotating (Kerr) black hole. These
authors adopted approximation of geometrical optics and presented a
detailed discussion of periodic variations of the observed
frequency-integrated flux. Origin of the variability is twofold: (i)
energy of individual photons is affected by the Doppler effect (the
effect of special relativity), and (ii) photon trajectories are
influenced by the gravitational field of the central black hole (purely
general-relativistic effect). Motivation to study relativistic effects
on light from a source near a black hole comes from astronomy, however,
no direct observational comparisons were possible in the early 1970s
when the formalism was first established.

During the last decade the interest of astronomers in modelling detailed
features in the light curve of an orbiting source has revived. Satellite
monitoring of active galactic nuclei (AGNs) focused our attention to
X-ray variability (Duschl, Wagner \& Camenzind 1991; Miller \& Wiita
1991) as a means of investigating the central object, presumably a
massive black hole surrounded by an accretion disk (Rees 1984).
Variability time-scales indicate that X-rays originate in the innermost
region of a few gravitational radii: $R\lesssim25\,R_g$ (e.g., Heeschen
\etal 1987). Power spectrum of AGNs has a particularly complex,
featureless behavior at frequency $\omega\gtrsim10^{-5}\,$Hz. The
fluctuating signal can be represented, in the frequency domain, by a
power-law: $F_{\omega}\propto\omega^{-\alpha}$ with
$1\lesssim\alpha\lesssim2$ (Lawrence \etal 1987). No strictly periodic
AGNs have been discovered but the power spectrum of some objects
(Papadakis \& Lawrence 1995) contains excessive power at frequency
$\omega\gtrsim10^{-2}\,$Hz, indicating quasi-periodic oscillations
(QPOs) analogous to those associated with low-mass X-ray binaries (Lewin
\& Tanaka 1995).

In spite of the evident progress on the observational side, theoretical
understanding of the short-term variability of AGNs remains rather
insufficient. Several physical mechanisms have been proposed, referring
to instabilities in accretion disks and associated jets. Possible
existence and properties of vortices on accretion disks have been
discussed by Abramowicz \etal (1992) and Adams \& Watkins (1995). It is
speculated that these vortices survive over several (perhaps many) local
orbital periods. Radiation from separate vortices is modulated by their
orbital motion around the black hole and a large number of individual
contributions results in observed fluctuations. In a more
phenomenological approach, one refers to spots (localized regions of
enhanced or diminished emissivity) residing on the surface of the disk
(Abramowicz \etal 1991; Wiita \etal 1991; Zhang \& Bao 1991). It is not
crucial for the form of the power spectrum whether the spots are
identified with vortices or whether they are of different physical
nature. Our present contribution deals with the bright-spot model but we
would like to emphasize that other scenarios for the short-term
variability of AGNs have been proposed. Krolik \etal (1991) explored
ultraviolet continuum variability and pointed out that independent
fluctuations originating at different radii in the source can produce
power-law spectra. Mineshige, Ouchi \& Nishimori (1994) employed the
theory of self-organized critical states that develop and persist in the
disk material, also leading to the observed power spectra. Kanetake,
Takeuti \& Fukue (1995) proposed that vertical oscillations of accretion
disks are the source of QPOs and studied characteristic periods of these
oscillations. Ipser (1994) and Perez \etal (1996) associated QPO
phenomena with frame-dragging effects which determine behavior of
relativistic disks, and in particular with low-frequency modes in
accretion disks near a Kerr black hole. Lamb \etal (1985), in the
context of Galactic QPO sources, explored interaction of a clumpy disk
with magnetosphere around a central object. At present, it is probably
fair to say that the relation between various possible instabilities and
the resulting variable signal of AGNs is, at the best, only vaguely
understood. It is thus impossible to discriminate between intrinsically
different models, even when they are testable in principle and the
necessary information is contained in the observed signal. In other
words, it appears encouraging that some features of the power spectra
can be explained by the bright-spot model but, in reality, this fact
does not mean very much when the emission properties of individual spots
are largely unknown. One can say (with Antoine de Saint-Exup\'ery) that
``Truth is not that which is demonstrable, but that which is
ineluctable.''

Until the theory of (nonaxisymmetric) instabilities in accretion disks
yields specific results about radiation outgoing from the disk surface,
phenomenological models must take all possible values of their parameter
space into account. This has not yet been possible to carry out with the
bright-spot model which has a number of degrees of freedom: distribution
of the spots across the disk, intrinsic emissivity properties of
individual spots, inclination of the observer, etc. Several numerical
codes evaluating the light-curve profiles and corresponding power
spectra have been developed (Asaoka 1989; Bao 1992; Karas, Vokrouhlickì
\& Polnarev 1992; Zakharov 1994) but models with a large number of
different spots still remain too expensive computationally (some steps
in the original derivation given by Cunningham \& Bardeen [1973] need to
be evaluated numerically). The authors had to impose additional
assumptions on the model without a sound physical reason (a
``canonical'' number of 100 spots is just an example of such
restriction). The main aim of the present contribution is to approximate
the light curve of an individual orbiting source of light by a practical
fitting-formula. It is required that the formula, while reflecting
relativistic effects in the light curve, can be implemented in a fast
code and substitute extensive ray tracing which is otherwise necessary.
It will be specified later what can be considered as ``practical'' and
what is a suitable approximation for our purpose.

The fitting-formula makes it possible to overcome unnatural restrictions
which are routinely imposed on the bright-spot model, and to cover the
parameter space of various models in a much more complete manner. The
fitting-formula is given in the next section.

\section{THE FITTING-FORMULA}

\subsection{Assumptions}
It is assumed that the source of light is located on a circular orbit in
the equatorial plane of a Kerr black hole (Misner, Thorne \& Wheeler
1973). Prograde orbits of the source (co-rotating with the black hole)
and direct-image trajectories of photons (not crossing equatorial plane)
are considered. Observer's position is specified, in Boyer-Lindquist
coordinates, by $R\rightarrow\infty,$ $\theta=\theta_0.$ (Azimuthal
coordinate is arbitrary due to axial symmetry.) Observed radiation flux
is determined by propagating photons along null geodesics from the
source to the observer, in accordance with approximation of geometrical
optics. It is straightforward to calculate the light curve of a
point-like source. The light curve of a finite-size source can be
obtained by integrating over the surface of the source. Calculation of
the light curves of finite-size sources contributes significantly to the
total computational cost of the bright-spot model and we wish to make
this step much faster. Technical details of our code which calculates
light-curve profiles were described by Karas \etal (1992).

Standard notation (e.g., Misner \etal 1973) will be adopted throughout
this work and geometrized units $(c=G=1)$ will be used; $M$ denotes the
mass of the central black hole, $a$ is its dimensionless
angular-momentum parameter. All lengths and times are made dimensionless
by expressing them in units of $M.$ Gravitational radius of the Kerr
black hole is expressed in terms of $a$: $R_g=1+(1-a^2)^{1/2},$ $0\leq
a\leq1.$

It turns out that the light-curve profile is very sensitive to observer
inclination, $\theta_0,$ and radius of the orbit of the source, $R_s.$
We will thus focus our attention to these two parameters. Rauch \&
Blandford (1994) studied light curves of point-like sources which show
extremely high peaks when the source crosses a caustic. Shape of the
caustic depends on angular momentum of the black hole and the light
curve is thus sensitive also to the value of $a.$ The case is different
when a finite-size source which covers the whole caustic is considered
(or, alternatively, temporal resolution of observation is lower than the
caustic crossing-time); the caustic is then unresolved and the
high-magnification spikes are smoothed down. We will consider only a
situation when the source does not cross the caustic or when its size,
$d,$ exceeds the cross-sectional size of the caustic, $s$. For
$R_s\gg1,$ equation (6) of Rauch \& Blandford (1994) gives an asymptotic
formula: $s\approx0.34\,R_s^{-1}a^2\sin^2\theta_0.$ In our calculations,
it was assumed that each spot radiates isotropically in its local
comoving frame and that the local emissivity decreases exponentially
with the distance from the center of the spot; we checked that the
computed profiles depend only weakly on these assumptions when the size
of the spot satisfies condition $s\ll d\ll R_s.$

It is only a variable component of the signal which is relevant for the
source fluctuations. Zero-level of our light curves will thus be set at
minimum of the observed signal. Variable component of the radiation flux
(counts per second) will be given in arbitrary units. This arbitrary
scaling of the profile contains sufficient information about the light
curve when radiating spots are all located at a constant radius.
Additional information about the absolute value of the maximum counting
rate is necessary when a distribution of spots at different radii is
considered. In this case one has to prescribe total flux from individual
spots and their geometrical shape as a function of radius. Since we pay
little attention to the model-dependent quantities, the absolute scaling
of the counting rate is not discussed in the present contribution
(except for an illustrative example mentioned in next section).

Analogously to the absolute value of the counting rate, also the phase
of the light curve is arbitrary. We scale the phase to interval
$0\leq\varphi\leq1$ and define the phase to be equal to 0.5 for the
maximum magnification due to the focusing effect. This definition means
that phase 0.5 corresponds to photons coming approximately (exactly when
$a=0)$ from the opposite side of the orbit, across rotation axis to the
observer. The phase $\varphi$ increases linearly with time, as measured
by a distant observer. Relation between $\varphi$ and the orbital phase
of the spot on the disk surface is complicated due to time-delay, but it
has been taken into account automatically by integrating individual
photon trajectories. Maximum Doppler enhancement of the observed
radiation corresponds to phase $\approx0.75.$ Dimensionless phase can be
easily converted to time interval (as measured by a distant observer)
when a rotation law of the spots is specified. For example, $\Delta
t=R_s^{3/2}+a$ is the period of the Keplerian circular orbit (as
measured by a distant observer) which is appropriate for a thin disk in
the equatorial plane.

\subsection{Numerical Method and Results}
A series of the light-curve profiles corresponding to a spot orbiting in
the range of distances $R_s\leq45\,R_g$ were computed. A model function
$F(\varphi,R_s,\theta_0;p_k)$ was specified, depending nonlinearly on
unknown parameters $p_k,$ $k=1,\ldots,8.$ ($F$ is the measured
frequency-integrated flux in arbitrary units.) The form of this function
has been chosen on the basis of asymptotic behavior and our experience
with extensive calculations of the profiles (Karas \etal 1992):
\begin{eqnarray} F & = & \left(\frac{p_3\cos\theta_0}{R_s}+
p_7\left(R_s-1\right)^{2/5}\right)
\Bigg[1+\sin\left[2\pi(\varphi+p_4\,R_s-\sfrac{3}{5})+
\sfrac{\pi}{2}\right]\Bigg]^z\cos^{-2/3}\theta_0        \nonumber \\ & +
& \left(p_1+p_6R_s^{1/3}\right)\,\exp\left[-p_2\left|\varphi-
\sfrac{1}{2}\right|^{9/5}\right]\,\cos^{-2}\theta_0, \label{fit}
\end{eqnarray} where $$z=p_5\cos^{1/2}\theta_0+p_8\cos^{3/2}\theta_0.$$
Parameters $p_k$ were determined by the Levenberg-Marquardt method
(nonlinear least-square fitting; Press \etal 1994). The range of fitting
was restricted to \beq 20\deg\leq\theta_0\leq80\deg,\qquad 3\,R_g\leq
R_s\leq45\,R_g, \eeq which is where relativistic effects on the light
curve are most profound. Table~1 gives the results of the fitting for
two values of the black-hole angular momentum: $a=0$ (a nonrotating,
Schwarzschild black hole), and $a=1$ (an extremely rotating Kerr black
hole). The $a$-dependence is only weak and it reflects a shift of
relative phases of the two peaks in the light curve---the Doppler and
the focusation peak (both peaks are simultaneously visible only if
$\theta_0\gtrsim70\deg,$ however). The shift is caused by the
frame-dragging effect which operates near rotating black holes. Figure~1
illustrates a typical form of the profiles normalized to the maximum
flux. The curves have been plotted according to equation (\ref{fit})
with parameters taken from Table~1.

Until now, only quantities which depend weakly on local properties of
individual spots were discussed. However, some applications are strongly
model-dependent, for example when radial distribution of spots or
eclipses of spots by a thick disk are taken into account (Bao \&
Stuchl\'{\i}k 1992; Karas \& Bao 1992; Mangalam \& Wiita 1993). Vortices
located across the surface of an accretion disk (as proposed by
Abramowicz \etal 1992) can act as spots in our model. Maximum flux from
spots at different radii must then be specified. Figure~2 illustrates
the maximum flux as a function of radius for our simple model of
isotropically radiating spots with constant size $d.$ Relative fluxes
from spots orbiting at different radii can be obtained by scaling the
normalized light curves from Fig.~1 by a corresponding value of the
maximum flux. It is evident from Fig.~2 that, at small radii,
gravitational redshift decreases the total flux when
$\theta_0\lesssim50\deg,$ while the Doppler and the lensing enhancement
dominate for $\theta_0\gtrsim50\fdg$ It is straightforward to obtain
graphs analogous to Fig.~2 also for other models of the spot local
emissivity.

\bigskip\bigskip

\section{CONCLUSION}
Variable signal from a source orbiting around a black hole is
approximated by the fitting-formula (\ref{fit}). This formula presents a
simple {\it model\/} of the corresponding exact expression which appears
too complex and inconvenient for numerous applications. Within the
above-discussed range of validity of the formula, the normalized
light-curve profile is nearly independent of the source shape.
Approximations which have been adopted turn out to be adequate in many
astrophysically relevant situations, e.g. in exploring direct images of
finite-size spots. This subject has been explored by numerous authors
but the fitting-formula offers much faster opportunity for studying
light-curve profiles. It is suggested that the formula is very practical
for modelling short-term featureless X-ray variability and
quasi-periodic oscillations in active galactic nuclei within framework
of the bright-spot model (work in progress).

\acknowledgments{This work is supported by the grants GACR 205/94/0504
and GACR 202/96/0206 in Czech Republic, and by Wenner-Gren Center
Foundation in Sweden.}

\begin{table}
\begin{center}
TABLE 1\vspace{3ex}\\ {\sc Parameters of
the Fitting-Formula}\vspace{2ex}
\begin{tabular}{ccccccccc} \hline\hline
$a$ & $p_1$ & $p_2$ & $p_3$ & $p_4$ & $p_5$ & $p_6$ & $p_7$  & $p_8$\\
\hline $0 $&$ 0.021696$&$ 190.7236 $&$ 0.3476 $&$-0.0018 $&$ 3.5106
$&$-3.6\times10^{-5}$&$0.0124$&$-0.0231$\\ $1 $&$ 0.024258$&$ 181.8421
$&$ 0.0958 $&$-0.0032 $&$ 4.7862
$&$-3.0\times10^{-5}$&$0.0109$&$-0.1527$\\ \hline
\end{tabular}
\end{center}
\end{table}


\bigskip

\hrule

\bigskip

FIGURE CAPTIONS (Figures available upon request from the author)

Figure~1---Illustration of the fitting-formula. The light-curve profile
normalized to the maximum flux (the counting rate) is shown by solid
lines for three cases: (a) $R_s=3\,R_g,$ $\theta_0=80\deg;$ (b)
$R_s=44\,R_g,$ $\theta_0=80\deg;$ (c) $R_s=44\,R_g,$ $\theta_0=20\fdg$
Shapes of the normalized curve depend on $a$ only weakly; here $a=1.$

Figure~2---Maximum flux (in arbitrary units) from a spot as a function
of dimensionless radius $R_s/R_g$ of the orbit; (a) $a=0,$ (b) $a=1.$
Observer inclination is indicated by three symbols in each of the
graphs: ``$\bullet$''\ldots$\theta_0=20\deg,$
``$+$''\ldots$\theta_0=50\deg,$ ``$*$''\ldots$\theta_0=80\fdg$

\end{document}